\documentstyle[floats,aps,epsf,prb]{revtex}

\begin{document}
\draft

\twocolumn[\hsize\textwidth\columnwidth\hsize\csname
@twocolumnfalse\endcsname

\preprint{January 14, 1996}
\title{Semiconductor effective charges from tight-binding theory}
\author{J. Bennetto and David Vanderbilt}
\address{Department of Physics and Astronomy, Rutgers University,
Piscataway, New Jersey 08855-0849}

\date{\today}
\maketitle

\begin{abstract}
We calculate the transverse effective charges of zincblende compound
semiconductors using Harrison's tight-binding model to describe
the electronic structure.  Our results, which are essentially exact
within the model, are found to be in much better agreement with
experiment than previous perturbation-theory estimates.
Efforts to improve the results by using more sophisticated variants
of the tight-binding model were actually less successful.  The
results underline the importance of including quantities that are
sensitive to the electronic wavefunctions, such as the effective
charges, in the fitting of tight-binding models.
\end{abstract}
\vskip2pc]

\narrowtext

The Born effective charges $e_T^*$, also known as transverse or
dynamic effective charges, are the fundamental quantities which
specify the leading coupling between lattice displacements and
electrostatic fields in insulators.\cite{pick}  In general the
effective charges are site-dependent tensors,
\begin{equation}
P_i=\sum_{l,j} e^{*(l)}_{T_{ij}}\,u^{(l)}_j \,+\,O(u^2) \;\;,
\end{equation}
where $P_i$ is the polarization in cartesian direction $i$, and
$u^{(l)}_j$ is the displacement of sublattice $l$ in cartesian
direction $j$.  However, for compound semiconductors of the
zincblende structure, which are the focus here, it is easily
shown that the effective charges are scalars, and are equal and
opposite for cation and anion; it is conventional to use the
positive cation effective charge to characterize a given compound.
The effective charges for a variety of zincblende semiconductors
have been computed using {\it ab-initio} density-functional
linear-response theory and have been found to agree very well
with experiment.\cite{degir,karch}  However, it is interesting to
inquire whether more approximate schemes can give a good accounting
of the effective charges in compound semiconductors. If so,
additional insight into the chemical and physical factors that
affect the $e_T^*$ might be obtained.

One particularly attractive and well-known approximate scheme is
the universal tight-binding model of Harrison.\cite{Ha81,Ha89}  It
provides a straitforward and computationally efficient approach to
calculating electronic properties of solids using a minimal
orthogonal $sp^3$ basis set, with the Hamiltonian limited to the
on-site and nearest neighbor terms.  The on-site elements
$\epsilon_s$ and $\epsilon_p$ are taken from calculated free-atom
term values, while the interatomic elements
($V_{ss\sigma}$, $V_{sp\sigma}$, $V_{pp\sigma}$ and $V_{pp\pi}$)
are taken to be species-independent ``universal'' constants times
the inverse square of the distance.  Given its simplicity,
the model is impressively successful in estimating many electronic
properties of a wide variety of materials.\cite{Ha81,Ha89}

It is thus natural to ask what the Harrison tight-binding model
would predict for the effective charges of the zincblende compound
semiconductors.  Oddly, this question does not appear to have been
answered previously.  The only previous work of which we are aware
made use of a two-center perturbation approximation to obtain
estimates of the effective charges.\cite{KiHa91,KiMuHa92} This
approach used an expedient division of the effective charge into
``static'' and ``transfer'' charge contributions, with the
interpretation of the latter being open to some
question.\cite{explan} The purpose of this Report is to present
essentially exact calculations of the transverse effective charges
computed for zincblende II-VI, III-V, and IV-IV semiconductor
compounds using the Harrison tight-binding parametrization.  While
the results could have been obtained using linear-response
techniques, we found it simpler to to compute the $e_T^*$'s instead
from finite differences, calculating directly the change in bulk
polarization from a small displacement of one sublattice using the
formulation of King-Smith and Vanderbilt.\cite{KsVa93} Our
calculations are both closer to the experimental values, and more
strongly correlated with them, than the previous reults.  However,
they are still significantly lower than experiment, and the
correlation is still not very good.  We also tried including
off-diagonal position matrix elements, and considered a modified
universal tight-binding model that was proposed to incorporate
non-orthogonality of the basis functions.\cite{ScHa86}
Unfortunately, both modifications were found to {\it worsen} the
results.

The details of our theoretical approach are as follows.  We
consider each zincblend compound at its experimental lattice
constant, with and without displacements of one sublattice along
the $\hat z$ direction by $\pm 0.0001$\AA.  The Bloch functions are
computed in the tight-binding representation using standard direct
matrix diagonalization on a mesh of $k$-points.  According to the
theory of Ref.\ \onlinecite{KsVa93}, the electronic contribution to
the polarization takes the form
\begin{equation}
{\bf P}_e = - \frac{ie}{(2\pi)^3} \sum_{n=1}^M{\int_{BZ}{
d{\bf k} \left \langle u_{n{\bf k}}\left| \nabla_{\bf k} \right|
u_{n{\bf k}} \right \rangle   }} \;\;,
\end{equation}
where the sum runs over occupied bands and the $u_{n{\bf k}}$
are the periodic parts of the Bloch wavefuctions,
\begin{equation}
   u_{n{\bf k}}({\bf r})=e^{-i{\bf k}\cdot{\bf r}}
\psi_{n{\bf k}}({\bf r}) \;\;.
\label{bloch}
\end{equation}
We are only interested in the $z$ components of $\bf P$ for the
distortions considered. After discretization in $k$-space, these
are given\cite{KsVa93} as
\begin{equation}
P_z = - \frac{2e(\Delta k)^2}{(2\pi)^3} \sum_{{\bf k}_\perp}
\phi({\bf k}_\perp).
\end{equation}
where ${\bf k}_\perp=(k_x,k_y)$ is discretized on a mesh of
spacing $\Delta k$, and the contribution from a string of
$J$ $k_z$-points takes the Berry-phase form\cite{Be84}
\begin{equation}
\phi({\bf k}_\perp ) = {\rm Im} \ln
\prod_{j=0}^{J-1}{\det \left \langle u_{m,{\bf k}_\perp,k_j} |
u_{n,{\bf k}_\perp,k_{j+1}} \right \rangle } \;\;.
\label{berry}
\end{equation}
Here the argument of the determinant is a $4\times4$ matrix
corresponding to the fact that $m$ and $n$ run over the four
occupied bands.  We typically use a discretization onto a
16$\times$16  mesh in ${\bf k}_\perp$ space, and extrapolate to
$J=\infty$ using strings of $J=32$ and $J=64$ $k_z$ points.  The
trivial ionic contribution to $P_z$ is added, and the value of
$e_T^*$ deduced by simple finite differences.

\begin{table}
\caption{The transverse charge $e_T^*$ for zincblende
semiconductors calculated at the experimental lattice spacing
$d$, compared with perturbation estimates of Kitamura and Harrison
(KH) and experimental values.  O-- and NO-- indicate orthogonal and
non-orthogonal tight-binding models respectively, while --O and
--OD refer to diagonal and off-diagonal representations of the
position operator.
\label{table1} }
\begin{tabular}{lrrrrrr}
       & $d$(\AA) & O--D & O--OD & NO--D &
       KH\tablenotemark[1]  & Expt.\tablenotemark[2] \\
   SiC &  1.88 &  1.97 &  2.20 &  1.84 &       &  2.57 \\
    BN &  1.57 &  1.24 &  0.96 &  1.01 &       &  2.47 \\
    BP &  1.97 & -0.09 & -0.18 & -0.23 &       &       \\
   BAs &  2.07 & -0.39 & -0.42 & -0.54 &       &       \\
   AlP &  2.36 &  1.92 &  1.61 &  1.64 &       &  2.28 \\
  AlAs &  2.43 &  1.75 &  1.50 &  1.50 &       &  2.30 \\
  AlSb &  2.66 &  1.48 &  1.22 &  1.32 &       &  1.93 \\
   GaP &  2.36 &  1.88 &  1.57 &  1.62 &  0.89 &  2.04 \\
  GaAs &  2.45 &  1.73 &  1.47 &  1.51 &  0.71 &  2.16 \\
  GaSb &  2.65 &  1.41 &  1.12 &  1.29 &  0.40 &  2.15 \\
   InP &  2.54 &  2.26 &  1.94 &  1.99 &  1.26 &  2.55 \\
  InAs &  2.61 &  2.11 &  1.85 &  1.86 &  1.07 &  2.53 \\
  InSb &  2.81 &  1.86 &  2.14 &       &  0.75 &  2.42 \\
   BeS &  2.10 &  1.61 &  1.08 &  0.71 &       &       \\
  BeSe &  2.20 &  1.56 &  1.04 &  0.71 &       &       \\
  BeTe &  2.40 &  1.51 &  0.96 &  0.57 &       &       \\
   ZnS &  2.34 &  1.89 &  1.46 &  0.53 &  1.25 &  2.15 \\
  ZnSe &  2.45 &  1.86 &  1.47 &  0.50 &  1.15 &  2.03 \\
  ZnTe &  2.64 &  2.05 &  2.50 &       &  0.98 &  2.00 \\
   CdS &  2.53 &  1.98 &  1.61 &  1.10 &       &  2.77 \\
  CdTe &  2.81 &  1.92 &  1.53 &  0.41 &  1.24 &  2.35 \\
\end{tabular}
\tablenotetext[1]{Ref.\ \onlinecite{KiHa91}.}
\tablenotetext[2]{Ref.\ \onlinecite{KiMuHa92}.}
\end{table}

Strictly speaking, the polarization $\bf P$ and effective charge
$e_T^*$ are not well-defined until the matrix elements of the
position operator are specified in the tight-binding basis.  In the
context of the above formulation, these position matrix elements
are needed for the conversion (\ref{bloch}) between the $u_{n{\bf
k}}$ and $\psi_{n{\bf k}}$.  The simplest ansatz is to assume that
the position operator is diagonal in the tight-binding
representation, with elements reflecting the coordinates of the
atoms.  However, such an ansatz is rather unphysical; it would
imply that the center of charge of an $sp$ hybrid on an atom would
lie exactly at the center of that atom, whereas in reality it would
be displaced toward the principal lobe of the hybrid.  We report
our results first for the simple ``diagonal'' ansatz.  Later, we
discuss the effects of trying to improve upon this ansatz, as well
as the effect of including the non-orthogonality in the model of
Ref.\ \onlinecite{ScHa86}.

\begin{figure}
\epsfxsize=2.8in
\centerline{\epsfbox{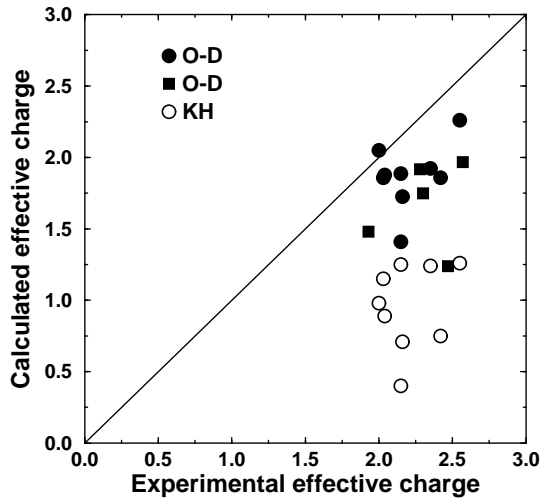}}
\medskip
\caption{Comparison of theoretical $e_T^*$ values from this work
(filled symbols) and from perturbation estimates of KH (open
symbols, from Ref.\ 6), plotted against experimental values.
Filled squares indicate results for compounds not considered by
KH.
\label{fig1}}
\end{figure}

The results for the orthogonal Harrison model\cite{Ha81} using the
diagonal representation of $\bf r$ are given in the column labeled
``O--D'' (orthogonal, diagonal) for a variety of zincblende
structures in Table I.\cite{explanwz}  The last two columns give
the values of the perturbation estimates of Kitamura and Harrison
(KH),\cite{KiHa91} and the experimental values, for comparison.
The data is also represented graphically in Fig.\ 1.  The filled
symbols are our results; the open ones are those given in Kitamura
and Harrison.\cite{KiHa91}  (The filled squares represent compounds
not studied in Ref.\onlinecite{KiHa91}.)  Our calculations shows a
clear improvement, although we still systematically underestimate
the experimental values of $e_T^*$.  The correlation between our
calculations and experiment is not very good, although it should be
noted that the lowest filled point is BN, a first row compound for
which the model is less accurate.

While the present results are certainly an improvement over the
perturbation estimates of KH, there is clearly room for
improvement.  We thus investigated two possible modifications of
the tight-binding model to see whether they would bring the
theoretical results into better agreement with experiment.  First,
we tried going beyond the artificial diagonal ansatz for the
tight-binding representation of the position operator by including
some off-diagonal terms.  Specifically, we included on-site matrix
elements between $s$ and $p$ orbitals, e.g., $\langle
s|z|p_z\rangle$.  The values of these matrix elements were obtained
from separate LDA calculations on free (neutral, spin-unpolarized)
atoms.  By symmetry, off-diagonal $p-p$ matrix elements of $\bf r$
are zero, and we assumed all off-diagonal intersite elements to be
zero as well.  The contribution of these extra off-diagonal terms
to the polarization $\bf P$ was calculated as a simple expectation
value, using the already-calculated wave functions (the Berry-phase
approach is not needed).  The results are given in the column
labeled ``O--OD'' (orthogonal, off-diagonal) in Table I, and are
compared with the previous results (open vs.\ closed circles) in
Fig.\ 2.  Unfortunately, the correction appears to be in the wrong
direction, and there is no apparent improvement in the correlation
between theoretical and experimental values.

\begin{figure}
\epsfxsize=2.8in
\centerline{\epsfbox{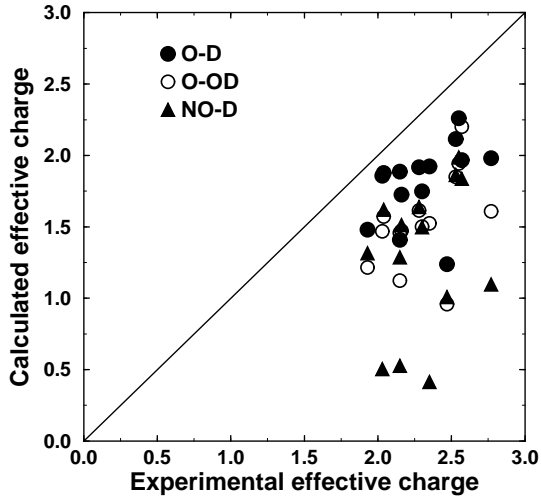}}
\medskip
\caption{Comparison of theoretical $e_T^*$ values using different
variants of the tight-binding model, plotted against experimental
values.  The notation is the same as in Table I.
\label{fig2}}
\end{figure}

Second, we attempted to improve the results by using a
tight-binding model that includes non-orthogonality of the basis,
as proposed by van Schilfgaarde and Har\-ri\-son.\cite{ScHa86} They
used extended H\"{u}ckel theory to derive the overlap elements
\begin{equation}
S_{ll'm} = \frac{2V_{ll'm}}{K(\epsilon_l + \epsilon_{l'})},
\end{equation}
from the original model, $\epsilon_l$ and $\epsilon_{l'}$ are the
onsite energies from the same model, and $K$ is a parameter
depending on row of the Periodic Table, chosen to fit the
equilibrium spacings of the IV--IV crystals.  The Hamiltonian
parameters were also renormalized following Eq.\ (11) of
Ref.\onlinecite{ScHa86}.  Some care is required in the application
of the theory of Ref.\onlinecite{KsVa93} to this case: the inner
product appearing in Eq.\ (\ref{berry}) has to be generalized to
take a form like $ \langle \phi_{m,{\bf k}_\perp,k_j} | S_{{\bf
k}_\perp,\bar k} | \phi_{n,{\bf k}_\perp,k_{j+1}} \rangle$, where
$|\phi_{n\bf k}\rangle$ is the vector of tight-binding coefficients
corresponding to $|u_{n\bf k}\rangle$, $S_{\bf k}$ is the overlap
matrix at wavevector $\bf k$, and $\bar k=(k_j+k_{j+1})/2$.  The
results are shown in the column labeled ``NO--D'' (non-orthogonal,
diagonal) in Table I, and as the filled triangles in Fig.\ 2.  Once
again, this ``correction'' is seen to act in the wrong direction,
worsening the agreement with experiment.

The failures of the above two attempts to improve the tight-binding
model are disappointing, but perhaps in hindsight they are not
surprising.  For the case of the non-orthogonal model, a partial
explanation may lie in the fact that the non-orthogonality was
added in large part to improve the fit for structures that were not
four-fold coordinated, which is not relevant here.  But more
fundamentally, we note that the model Hamiltonians we tested were
developed by fitting to energy bands; thus, the fit included only
information about energy eigenvalues, and not the wavefunctions per
se.  However, the electric polarization is a quantity which depends
sensitively on the electronic wavefunctions themselves.  Thus, a
real improvement in the tight-binding model can probably best be
accomplished by including quantities that are sensitive to the
wavefunctions, such as $e_T^*$ values, in the fitting procedure
itself.

In summary, we have carried out essentially exact calculations of
the transverse effective charge $e_T^*$ in compound semiconductions
within Harrison's universal tight-binding scheme.  We find a
significantly improved agreement with experiment, compared with
previous perturbation estimates.  However, the theoretical results
still show a systematic underestimate relative to experiment, by an
average of 20\%.  Attempts to improve the agreement by including
off-diagonal position matrix elements, or non-orthogonality of the
basis, were actually found to lead to a worsening agreement with
experiment.  Based on this experience, we suggest that it might be
helpful to use the effective charge as a fitting parameter in
future tight-binding models.  Such an approach might lead to a more
accurate description of the electronic properties of semiconductors
within this class of simple, but very useful, models.

This work was supported by ONR Grant N00014-91-J-1184.
J.B.\ acknowledges support of ONR Grant N00014-93-I-1097.


\begin{references}

\bibitem{pick} R.M. Pick, M.H. Cohen, and R.M. Martin, Phys. Rev. B
{\bf 1}, 910 (1970).

\bibitem{degir} S. de Gironcoli, S. Baroni, and R. Resta,
Phys. Rev. Lett. {\bf 62}, 2853 (1989).

\bibitem{karch} K. Karch {\it et al.}, Phys. Rev. B. {\bf 50},
17054 (1994).

\bibitem{Ha81} W.A.~Harrison, Phys. Rev. B {\bf 24}, 5835 (1981).

\bibitem{Ha89} W.A.~Harrison, {\it Electronic Structure and the
Properties of Solids} (Freeman, San Francisco, 1980) (reprinted by
Dover, New York, 1988).

\bibitem{KiHa91} M.~Kitamura and W.A.~Harrison, Phys. Rev. B
{\bf44} 7941 (1991).

\bibitem{KiMuHa92} M.~Kitamura, S.~Muramatsu, and W.A.~Harrison,
Phys. Rev. B {\bf 46} 1351 (1992).

\bibitem{explan} In Refs.\onlinecite{KiHa91} and
\onlinecite{KiMuHa92}, the polarization change is taken to have
contributions from a ``static charge'' and a ``transfer charge.''
The former contribution is just the static charge residing on an
atom times the distance it moves, while the latter is taken as the
induced change in static charge times the interatomic separation.
This last approximation is especially questionable; even within the
stated assumption that electrons transfer only to neighboring
atoms, it neglects the possibility of charge transfer along bonds
other than the one on which the perturbation analysis is being
performed.

\bibitem{KsVa93} R.D.~King-Smith and D.~Vanderbilt, Phys. Rev. B
{\bf 47} 1651 (1993).

\bibitem{ScHa86} M.~van Schilfgaarde and W.~Harrison, Phys. Rev. B
{\bf 33} 2653 (1986).

\bibitem{Be84} M.V.~Berry, Proc. Roy. Soc. London A
{\bf 392} 451 (1984).

\bibitem{explanwz} We also calculated the $e_T^*$ tensor for the
same componds in the ideal wurtzite stucture, and found that
neither of the two independent $e_T^*$ components typically differs
by more than $\sim$2\% from the zincblende values.

\end{references}
\end{document}